# Electromagnetic Design of $\beta_g$= 0.9, 650 MHz Superconducting Radiofrequency Cavity

Arup Ratan Jana[1], Vinit Kumar[1], Abhay Kumar[2] and Rahul Gaur[1]

[1]Materials and Advanced Accelerator Science Division

[2]Power Supplies and Industrial Accelerator Division

Raja Ramanna Center for Advanced Technology, Indore, India

E-mail: arjana@rrcat.gov.in, arup.jana@gmail.com

**Abstract:** We present the electromagnetic design study of a multi cell, $\beta_g$= 0.9, 650 MHz elliptic superconducting radiofrequency cavity, which can be used for accelerating H$^-$ particles in the linear accelerator part of a Spallation Neutron source. The design has been optimized for maximum achievable acceleration gradient by varying the geometry parameters of the cavity, for which a simple and general procedure is evolved that we describe in the paper. For the optimized geometry, we have studied the higher order modes supported by the cavity, and the threshold current for the excitation of the regenerative beam break up instability due to dipole modes has been estimated. Lorentz force detuning studies have also been performed for the optimized design and the calculations are presented to find the optimum location of the stiffener ring to compensate for the Lorentz force detuning.

## 1. INTRODUCTION

Superconducting Radiofrequency (SRF) cavities have become an essential component of high energy, high power particle accelerators in modern times, owing to its several advantages compared to normal conducting cavities such as less power dissipation on the cavity wall, and the possibility of operating at larger beam aperture radius that allows higher beam current to be accelerated [1,2]. One of the most contemporary activity of exploring these advantages through the designing of an SRF cavity , as described in Ref. [3], tells us about 1.3 GHz TESLA type cavity which will be used in the squeezed International Linear Collider (ILC) section of the proposed high intensity H⁻ ion linac at Fermilab. The another important recent applications of SRF cavities is in building a spallation neutron source (SNS) at Oak Ridge National Laboratory, USA, which requires a negative hydrogen ion (H⁻) accelerator giving ~ 1 GeV energy with few mA pulsed current in ~ ms pulses at around 50 Hz [4]. The medium and high energy section (>200 MeV) of such an accelerator uses elliptic SRF cavities. There is a plan to build such an SNS in India for various applications. The linear accelerator for this proposed Indian Spallation Neutron Source (ISNS) will use two sets of multi cell superconducting elliptic cavities– one set for medium energy that will accelerate the beam from 200 MeV to around 500 MeV, and the other set for high energy that will accelerate H⁻ beam from 500 MeV to 1 GeV.  These multi cell cavities are operated in the π mode since it offers the maximum shunt impedance for a standing wave configuration. In the π mode, for synchronization between the beam and the RF, the cell length along the beam axis should be $\beta\lambda/2$, where $\beta$ is the velocity of H⁻ in unit of speed of light and $\lambda$ is the free space wavelength of the radiofrequency wave used for acceleration. Ideally, the cell length should be varying continuously as the beam traverses from one multi cell cavity to another, and gets accelerated.  However, in order to reduce the complexity of the manufacturing process, one goes for a limited number of sets of these multi cell cavities (which is two in our case) and the cell length is fixed in each set and is denoted by $\beta_g\lambda/2$, where $\beta_g$ is called the geometric beta. For the medium energy section, we have chosen $\beta_g$ = 0.61, and for the high energy section, we have chosen $\beta_g$= 0.9.

In this paper, we present the design study of a 650 MHz, multi cell elliptic SRF cavity for $\beta_g$= 0.9 section. We have performed a detailed parametric study to optimize various geometric dimensions of the cavity, assuming the standard shapes constructed using two elliptic arcs as shown in figure1, which has evolved over several years to minimize multipacting problems[1,2,5]. It is interesting to point out that during the evolution of a suitable cavity shape to avoid multipacting, several other shapes, e.g., spherical shape [6,7], were also proposed. The spherical shape however had the disadvantage that the axial length of the cavity was constrained to be equal to the diameter that was fixed for a particular value of operating frequency and thus there was no flexibility in optimizing the cavity length for synchronizing with particle of a given velocity. The shape involving two elliptic arcs, which we described above, does not have these problems and has become standard choice for designing superconducting cavity for medium and high energy accelerators, which we have opted here. For this shape, we have done the geometry optimization ,mainly for achieving maximum possible acceleration field $E_{acc}$ in the cavity. The maximum possible value of accelerating field $E_{acc}$ in the cavity is limited by the upper limit on the peak surface magnetic field $B_{pk}$ that leads to breakdown in the superconducting properties and the peak surface field $E_{pk}$ that leads to field emission. The optimized geometry has therefore been evolved in our design that minimizes $B_{pk}/E_{acc}$, keeping a satisfactory value for $E_{pk}/E_{acc}$. This criterion for optimization is similar to the one adopted for the design of $\beta_g =1$, 1300 MHz cavities in the literature[8]. The dimensions of mid-cell geometry have been optimized for achieving maximum acceleration gradient, for which a simple and general procedure is evolved in this paper. Belomestnykh [5] and Shemelin [9] have discussed a procedure for the optimization of the cavity geometry, which is essentially a multi dimensional optimization. In contrast, in the procedure that we have developed, we perform step by step, one dimensional optimization, which is relatively simple. After optimizing the mid cell geometry, the end-cell geometry is optimized, keeping the requirement on field flatness in the cavity [9,10]. A two-dimensional electromagnetic design code SUPERFISH [11] has been used for performing these calculations.

While performing these optimization studies, we have explored the effect of cavity wall angle. The cavity wall angle is defined as the angle made by the common tangent to the iris and equator ellipses with the beam axis as shown in figure 1. Note that in this convention, if the wall angle is greater than $90^0$, the cavity geometry is called the re-entrant geometry. Generally, there is a constraint on the maximum value of the wall angle $\alpha$, which is decided by the mechanical considerations, ease of chemical cleaning [12], *etc*. In our optimization studies, we have explored the case when there is no such constraint on $\alpha$, and found that the performance is improved for higher values $\alpha$. We have therefore chosen maximum possible value of $\alpha$ that is practically possible and used this as a constraint in the study of optimization of mid cell geometry.

An important aspect in the design of SRF cavity is the study of Higher Order Modes (HOMs) [1,3,13],because it decides the maximum current that can be accelerated without exciting unwanted beam instability. We have studied the HOMs that can be excited in the cavity, and calculated the shunt impedance and the quality factor of prominent HOMs using the electromagnetic design code SLANS [14]. The wake loss factor has been evaluated using an electromagnetic design code ABCI [15]. This study is useful in assessing the suitability for operation at the desired beam currents.

Another important design aspect of such SRF cavities is the study of Lorentz Force Detuning (LFD) [16], where the detuning in the cavity frequency after the application of RF power is studied. In order to keep the detuning within an acceptable level, the thickness of the cavity wall is chosen suitably and also the provision is made for stiffening rings. We have performed this study using the finite element code ANSYS [17].

The paper is organized as follows. Section 2 describes the studies for choosing the optimum geometry for the mid cell. The first subsection discusses the rationale for the choice of different design parameters, like frequency, beam aperture radius, *etc*., and the general features of the optimization process. The second subsection discusses the details of design calculations performed for the optimization. Design of the multi cell cavity is described in Section 3, where we first discuss the criteria for choosing the number of cells in the cavity and then the geometry of the end cell is optimized. Section 4 describes the studies of the HOMs supported by the cavity, where the calculations of frequencies and shunt impedances of HOMs are presented. Calculation of wake loss parameter is also discussed there. The details of calculations performed for LFD studies is described in Section 5. Finally, we present some discussions and conclude in Section 6.

## 2. OPTIMIZATION OF THE MID CELL GEOMETRY

*2.1 Generalities.* The first issue that needs to be settled before starting the design is the operating frequency. For superconducting cavity based on Niobium operating at 2K, as described in Ref. [1], a frequency range 0.3 – 3 GHz is suitable from the point of view of minimum power loss on the cavity wall surface. The higher end in this frequency range is preferred for low beam current application and the lower end is preferred for high beam current application. This is because for higher frequency, the aperture available for the transmission of beam is small, and therefore the threshold current for the excitation of beam instability will be low. Since the average beam current for SNS applications is typically ~ 1 mA, the lower frequency in the range 0.3 – 3 GHz is preferred. The frequency for the SNS at Oak Ridge National Laboratory has been chosen to be 805 MHz [4]. In the case of ISNS, it is being planned to use solid state RF power source at 650 MHz. The operating frequency is therefore chosen as 650 MHz for the SRF elliptic cavities.

After having chosen the operating frequency, we discuss about the optimization of the geometry of the elliptic cell of the SRF cavity. As will be discussed later, for the multi cell cavity, the end cells have slightly different geometry compared to the inner cells. In this section, we describe the geometry optimization of inner cells, which are identical to each other and we call them as mid cells. We are following a typical TESLA type cavity shape [2], based on two elliptic arcs joined by a straight line, as shown in figure 1. The three-dimensional cavity shape is a figure of revolution around the beam axis, obtained using the contour shown in figure1, and is described by the following seven independent parameters– iris ellipse radii *a* and *b*, equator ellipse radii *A* and *B*, iris radius $R_{iris}$, equator radius $R_{eq}$, and half-cell length *L*. The wall angle $\alpha$ shown in the figure is a parameter that can be derived from these seven parameters. The correlation between the cavity electromagnetic performance and its geometry is summarized below:

- As discussed earlier, since the $TM_{010}$ - $\pi$ mode is chosen as the operating mode of the cavity, the cell length should be equal to $\beta_g\lambda/2$ for synchronization. Half cell length $L = \beta_g\lambda/4$ is therefore fixed in our design and is equal to 103.774 mm.

- $R_{iris}$ is governed by the beam dynamics considerations as well as by the requirement of the cell-to-cell coupling, designated by $k_c$. Higher value of $R_{iris}$ reduces the effect of wake field and HOMs, and gives higher value of $k_c$. However, it reduces the shunt impedance of the fundamental accelerating mode, and also reduces the maximum achievable accelerating gradient. We have chosen $R_{iris}$ = 50 mm in our design and later show in the paper that this choice is acceptable for the design value of beam current. In this paper, we have kept $R_{iris}$ fixed and have not optimized its value.

- The slope $\alpha$ and the location of the common tangent shown in figure 1 influence the ratio of the volumes of different cavity regions storing electric and magnetic energy. This is because equatorial region prominently stores magnetic energy and iris region prominently stores electric energy. Apart from having a minor effect on cell-to-cell coupling, the wall angle $\alpha$ therefore play an important role in deciding the electric and magnetic peak fields of the cavity.

- The ratio $a/b$ influences the ratio $E_{pk}/E_{acc}$. We therefore optimize $a/b$ such that $E_{pk}/E_{acc}$ is minimum in order to achieve maximum acceleration gradient.

- The ratio $A/B$ is purely decided by mechanical requirements like stiffness, rigidity, *etc*., of the cavity and has almost no influence on the cavity electromagnetic performance.

- $R_{eq}$ is used for the frequency tuning and has almost no effect on the electromagnetic characteristics of the cavity as well as its mechanical properties.

We thus find that out of the seven independent parameters that describe the geometry in figure 1, two parameters, i.e., $R_{iris}$ and $L$ are fixed. Also, for each simulation reported in this paper, $R_{eq}$ is tuned to achieve the resonant frequency. We are thus left with four independent parameters to consider, while optimizing the geometry. In the existing literature [5,9], there are two types of cavity performance optimizations that have been discussed – (i) optimization for the maximum acceleration gradient $E_{acc}$, such that the accelerator can be made as compact as possible. This optimizes the construction cost of the accelerator, (ii) optimization for the minimum power loss on cavity wall such that there is less cryogenic load to be handled. This optimizes the operational cost. We discuss the issues involved in each of these in the following paragraphs.

First, we discuss the issues involved in the optimization for the maximum acceleration gradient. In a SRF cavity, $E_{acc}$ is limited by the requirement that the peak magnetic field $B_{pk}$ developed on the inner surface of the cavity wall must be less than the critical magnetic field. This limit is popularly known as the hard limit [1]. For 650 MHz, a safe value for $B_{pk}$ is typically taken as 70 mT [18] . Also, as mentioned earlier, the peak electric field $E_{pk}$ on the cavity surface can cause field emission and should be below a certain critical value. Maximum tolerable value of $E_{pk}$ can however be increased by improving the surface finish, cleanliness and by following an appropriate surface processing technique. This limit is therefore called the soft limit[1]. For 650 MHz, a safe value for $E_{pk}$ is typically taken as 40 MV/m [19].

Second, we discuss the issues involved in the optimization for the minimum power loss on cavity wall $P_c$. In order to minimize $P_c$, we maximize a parameter called $GR/Q_0$, which is evident from following expression [1,5],

$$P_c = \frac{V_{cav}^2}{R} = \frac{V_{cav}^2 \cdot R_s}{G(R/Q_0)} \tag{1}$$

Here $G = R_s Q_0$ is the geometry factor that is independent of cavity material and only dependent on the cavity geometry, $R_s$ is the surface resistance of the material of the cavity, $Q_0$ and $R$ are respectively the unloaded quality factor and the shunt impedance of the cavity, and $V_{cav}$ is the accelerating voltage.

Either of the two approaches mentioned above have been used in the literature [5,9] for the optimization of mid cell geometry. In this paper, we have optimized our design for maximum acceleration gradient, but we finally show that it is moderately optimized even for minimum power loss. In a resonating cavity, the peak electric and magnetic field typically has an inverse relation, i.e., if we perturb the cavity geometry to increase the peak magnetic field, the peak electric field decreases. Therefore, one can achieve a higher accelerating gradient for a relatively low peak surface magnetic field if one is ready to tolerate more peak electric field and for a relatively low peak surface electric field if one is ready to tolerate high peak magnetic field. Hence, in the approach that we have followed in this paper, we have tried to optimize the geometry to minimize the value of $B_{pk}/E_{acc}$ for a *fixed* value of $E_{pk}/E_{acc}$.

To summarize, one needs to find optimum mid cell geometry to (i) maximize $GR/Q_0$ for minimizing the heat loss on cavity walls, and (ii) minimize $B_{pk}/E_{acc}$ for a suitably chosen $E_{pk}/E_{acc}$ for maximizing the achievable acceleration gradient.

*2.2 Design Optimization Calculations.* We now discuss our procedure for the optimization of the geometry of the mid cell. We keep a target value of $E_{pk}/E_{acc} \leq 2$ and $B_{pk}/E_{acc} \leq 3.78$ mT/(MV/m) such that an accelerating gradient up to ~ 18.5 MV/m can be achieved. As discussed earlier, we have four independent variables in the cavity geometry – *a*, *b*, *A*, and *B*. We start with the simplest possible geometry, where $a/b = A/B = 1$ and $\alpha = 90^0$. With these three constraints, there will remain only one independent variable. Figure 2 shows the plot of $B_{pk}/E_{acc}$ and $E_{pk}/E_{acc}$ as a function of *B*, which is taken as the independent variable. Note that all the simulations presented are performed using SUPERFISH. Calculations were done, starting from $B = L/2$. We observe that initially, $B_p/E_{acc}$ is very high [~ 5 mT/(MV/m)], whereas the value of $E_{pk}/E_{acc}$ is moderate (~ 2.3). As we increase $B$ and $A$ together, this enhances the magnetic volume of the cavity and the ratio $B_{pk}/E_{acc}$ decreases steadily. Up to around $B = 84$ mm, this happens with very insignificant increment in the value of $E_{pk}/E_{acc}$. After this point, although the value of $B_{pk}/E_{acc}$ keeps on decreasing slowly, the increase in the value of $E_{pk}/E_{acc}$ becomes very significant. Therefore it is reasonable if we stop at the value of $A=B= 84.00$ mm (and $a=b=19.774$mm) and choose this geometry for the further optimization. Here, for this geometry the values of $E_{pk}/E_{acc}$ and $B_{pk}/E_{acc}$ are ~2.3 and ~3.7 mT/(MV/m) respectively. These values are approaching close to our targeted values. Based on these simulations, we fix the value of $B$ as 84 mm and perform the optimization study of remaining geometry parameters.

We would like to mention here that, although an initial optimum value has been obtained for four variables - $a$, $b$, $A$ and $B$, we have particularly chosen $B$ as the parameter whose value is frozen, because we observe that the value of $B$ affects the electromagnetic property of the cavity the least. This is perhaps because even when we change $B$, the value of $R_{eq}$ that is used to tune the cavity frequency does not change much and therefore the overall cavity geometry is not significantly affected by the change in $B$.

In the preceding discussions, we have taken $\alpha = 90^0$. It turns out that although the RF performance of the cavity is expected to improve with the increasing wall angle as will be shown later in this section, the mechanical and other consideration put a limit on $\alpha$. For the further optimization of the mid cell geometry, we have therefore kept the wall angle fixed at $85^0$. We have checked that the result mentioned in the previous paragraph regarding the optimum value of $B$ does not change for the case $\alpha = 85^0$. Second, as discussed earlier, for the further optimization, we have decided to keep $E_{pk}/E_{acc}$ fixed at 2.0, and minimize $B_{pk}/E_{acc}$ with this constraint. Having fixed the value of $B$, and having a constraint that $\alpha = 85^0$, we are left with only two independent variables in the geometry of the mid cell. Further, using the constraint that $E_{pk}/E_{acc} = 2.0$, we are left with only one independent variable, which we use to minimize the value of $B_{pk}/E_{acc}$ to fix the mid cell geometry.

Figure 3 shows the variation of $B_{pk}/E_{acc}$ as a function of $A$ for different values of $a/b$, keeping $\alpha = 85^0$ and $B = 84$ mm. Since there are three additional constraints now (on $\alpha$, $B$ and $a/b$), $A$ is the only independent variable for each plot shown in figure 3. We observe that $B_{pk}/E_{acc}$ keeps monotonically decreasing with $A$. This happens because as we increase the value of $A$, the equatorial volume increases, as a result of which, the peak magnetic field decreases.

In figure 4, we have plotted the corresponding value of $E_{pk}/E_{acc}$ as a function of $A$ for the same set of calculations, and we observe that $E_{pk}/E_{acc}$ keeps on monotonically increasing with $A$. We then keep a constraint that we restrict the value of $E_{pk}/E_{acc}$ to less than or equal to 2.0. With this constraint, as seen in figure 4, the maximum possible value of $A$ is around 83.00 mm. We want to keep the value of $A$ as large as possible because $B_{pk}/E_{acc}$ reduces as $A$ is increased. We make an interesting observation when we plot $E_{pk}/E_{acc}$ as a function of $a/b$ for different values of $A$ in figure 5. We notice that for each value of $A$, there is an optimum value of $a/b$ for which $E_{pk}/E_{acc}$ is minimum. This justifies the importance of choosing $a/b$ as an optimization parameter. The reason for the existence of such a minimum is that the distribution of the electric field on the cavity surface ($E_s$) is nearly uniform near the iris region when this minimum occurs. This is explicitly shown in figure 6, where $E_s/E_{acc}$ on the cavity surface is plotted in a region near the iris as a function of distance $z$ along the beam axis ($z = 0$ corresponds to the equatorial plane).

The summary of the observations made so far is that although a higher value of $A$ helps in reducing $B_{pk}/E_{acc}$, it increases $E_{pk}/E_{acc}$. However, for each value of $A$, there exists an optimum value of $a/b$ for which $E_{pk}/E_{acc}$ is minimum. As discussed earlier, our target value of $E_{pk}/E_{acc}$ is 2. Based on the data shown in figure 3 and figure 4, we can choose a range of a/b ~ 0.55 - 0.6, and $A$ ~ 83 for the further fine tuning of the optimization. Figure 7 and figure 8 show plots of $B_{pk}/E_{acc}$ and $E_{pk}/E_{acc}$ respectively as a function of $A$ for different values of $a/b$ within a narrow range compared to figure 3 and figure 4. From the data presented in figure 8, we choose the points corresponding to $E_{pk}/E_{acc}$ = 2.0. These points are denoted by A, B, C, D,E and F in figure 8. Next, we find out the value of *A* for each of these points from figure 8 and then use figure 7 to find out corresponding value of $B_{pk}/E_{acc}$ for each of these points and plot them in figure 9a and figure 9b. We emphasize that for the plots shown in these figures, $E_{pk}/E_{acc}$ = 2. From these two plots, we conclude that the optimum value of *A* and *a/b* respectively that minimizes $B_{pk}/E_{acc}$ are 83.275 mm and 0.57 respectively.

The final parameters describing the optimized geometry of the mid cell are listed in table 1. The RF calculations for the π-mode have been done for this geometry using SUPERFISH and the results are tabulated in table 2. Figure 10 shows the field contours for the mid cell geometry and $TM_{010}$ like mode is clearly seen.

Finally, we would like to discuss the effect of cavity wall angle *α* on the electromagnetic performance of the cavity. The calculations presented so far in this section assume that $\alpha = 85^0$. We have performed the complete set of calculations described so far for different values of α. Figure 11 shows a plot of the most optimum value of $B_{pk}/E_{acc}$ achieved for different values of *α*. We notice that it is possible to reduce the minimum achievable value of $B_{pk}/E_{acc}$ if we go for higher value of *α*. This is perhaps because as we increase *α*, the volume of the equatorial region of the cavity increases, and therefore the peak magnetic field decreases. However, from the manufacturing point of view, larger is the wall slope angle, more is the inconvenience during the cavity fabrication. Also, in this case it becomes more difficult to drain the residual liquids from the multi cell geometry after chemical cleaning and other processes. Considering these constraints, we have kept *α* fixed at $85^0$. However, if these constraints are not present, higher wall angle is preferred.

In the same figure, we have also shown the plot of $GR/Q_0$ as a function of *α* and we observe that it shows a tendency to saturate for higher wall angle. This indicates that from the electromagnetic point of view, both for lower loss and higher accelerating gradient, a re-entrant geometry is a better choice. This is in agreement with the observations reported in Ref. [9] that the re-entrant geometry is better for achieving higher acceleration gradient as well as minimum cavity loss.

**3. OPTIMIZATION OF PARAMETERS FOR MULTI CELL CAVITY**

In the previous section, we have discussed about optimization of the geometry of a single cell. In real practice, multi cell cavity is used in particle accelerators because it is more efficient and provides higher value of effective acceleration gradient. The first issue that needs to be decided is the number of cells $N$ in the cavity. There are a number of considerations that go in deciding the number of cells. First consideration is the transit time factor $T$ that affects the energy gain $\delta W$ per cell by the Panofsky formula given by $\delta W = qE_0 Tl \cos\varphi$ [13], where $q$ and $E_0$ stand for the charge of the particle and the average of the electric field amplitude along the beam direction, $l$ is the cell length and $\varphi$ is the phase of the particle. The cavity will have the maximum value of $T$ for $\beta = \beta_g$, for which it is designed. As the particle gets accelerated, its $\beta$ keeps changing, and the dependence of $T$ on $\beta$ is given by the following equation[20]:

$$T(N, \beta, \beta_g) = \left(\frac{\beta}{\beta_g}\right)^2 \cos\left(\frac{\pi N}{2\beta/\beta_g}\right) \frac{(-1)^{(N-1)/2}}{N\left((\beta/\beta_{Gg})^2 - 1\right)}, \quad \text{for } N = \text{odd}$$

$$= \left(\frac{\beta}{\beta_g}\right)^2 \sin\left(\frac{\pi N}{2\beta/\beta_g}\right) \frac{(-1)^{(N+2)/2}}{N\left((\beta/\beta_g)^2 - 1\right)}. \quad \text{for } N = \text{even} \quad (2)$$

Figure 12 shows the dependency of the transit time factor $T$ on the normalized particle velocity ($\beta/\beta_g$) for different values of $N$. It is seen that as the number of cells in a cavity increases, the transit time factor falls sharply with deviation in $\beta$ from $\beta_g$. This means that the range over which such a multi cell cavity will work gets restricted. We are planning to use these cavities for accelerating H⁻ particles over an energy range 500 MeV – 1 GeV. This corresponds to a range in $\beta/\beta_g$ from 0.84 to 0.97. For different values of the number of cells, the range in the value of $T$ corresponding to this range in $\beta/\beta_g$ can be seen in figure 12. We observe that beyond $N = 5$, the transit time starts going below 0.5. We have thus chosen $N = 5$.

We have also looked at this issue from the point of view of field flatness $\eta$ in the multi cell cavity, which is defined as[21]

$$\eta = 1 - \frac{E_{max} - E_{min}}{\frac{1}{N}\sum_1^N (E_i)}, \qquad (3)$$

where $E_i$ is the electric field amplitude in $i^{th}$ cell and $E_{max}$ and $E_{min}$ denote respectively the maximum and minimum of these amplitudes. A value of $\eta$ close to 1 is desired to maintain a good synchronism of the accelerating particle with the electromagnetic field. The field flatness is related to the difference between the resonant frequency of two successive modes and the relation can be expressed as follows [1],

$$\eta \sim 1 - \frac{\omega_n - \omega_{n-1}}{\omega_{avg}} \cong 1 - k_c\left(1 - \cos\frac{\pi}{N}\right) \cong 1 - 2k_c \sin^2\left(\frac{\pi}{2N}\right) \cong 1 - \frac{k_c}{2}\left(\frac{\pi}{N}\right)^2, \qquad (4)$$

where $\omega_n$ is the resonant frequency of the $n^{th}$ structure mode and $\omega_{avg}$ is the average value. We have evaluated the value of $k_c$ using the results of single cell simulation. The difference between the resonating frequency $f_\pi$ of the $\pi$ mode and $f_o$ of the $0$ mode gives the bandwidth and this can be used to calculate $k_c$ as illustrated in table 3, where we get $k_c = 0.8025\%$. With this value of $k_c$ and $N=5$, we get field flatness as 0.9922, which is similar to the field flatness obtained by plugging in the number of TESLA design [2], where $k_c = 1.87\%$ and $N = 9$.

Another consideration in choosing the number of cells in the multi cell cavity is the availability of the RF power source with suitable capacity and also the power handling capacity of the input power coupler. Here, for a 5-cell cavity, the power taken by beam will take around 75 kW. The commonly used couplers for the SRF cavities can handle a power of 100 kW easily[22].

Next, we discuss about the geometry of the end cells. For the optimization of the end cells, one has to follow a different approach compared to the mid cell. The reason for this is that the end cell is connected to the mid cell at one end and to the beam pipe on the other end. Thus, unlike the mid cell, the end cell does not see the symmetric boundary condition at the two ends. Due to this asymmetry, even in the π mode, the electric field terminates at the two ends of the end cell differently. This may therefore result in a slightly different resonant frequency of the multi cell cavity compared to the mid cell frequency if the end cell geometry is taken exactly the same as mid cell geometry. Also, this results in an unequal accelerating gradient in the consecutive accelerating cells. Restoring the resonant frequency and the field flatness therefore becomes an important part of the end cell optimization. Thus a slight modification in the geometry of the end cell is needed to restore the cavity resonant frequency to that of the resonating frequency of a single mid cell and to improve the field flatness. One possible approach to tune the cavity to get back the mid-cell frequency is to change the length of the end cell. The advantage here is that one can keep the equator radius of the end cells equal to that of the mid cell, which is desired for the ease of fabrication.

We now present the results of our calculation, where we have studied the effect of variation in the length of the end cell on field flatness and the resonant frequency. The end cell can be imagined to be divided into two half cells about the equator plane and here we have kept the geometry of the half cell adjacent to the mid cell the same as that of the mid cell. We have only varied the length $L_e$ of the half cell adjacent to the beam pipe. Figure 13 shows the dependence of the resonant frequency and the field flatness on $L_e$. The resonant frequency monotonically increases with $L_e$, but the field flatness shows a maximum when the total cavity frequency is restored back to the resonant frequency of a single mid cell. We keep the value of $L_e$ corresponding to this maximum as the design value, which is 105.8 mm. The value of wall angle corresponding to this is $82.75^0$.

After having chosen *N* and optimized the end cell geometry, we have the full cavity geometry. We now present the results of the electromagnetic field calculations of the 5-cell cavity. Figure 14 shows the profile of the magnitude of the axial electric field in the cavity. Field contour plot in one half of the multi cell cavity is shown in figure 15. For a 5-cell cavity, we expect five normal modes and the resonant frequency of different modes as a function of mode number are plotted in figure 16. Using this plot, we have calculated $k_c$ = 0.751%. Note that this is slightly different from the earlier value discussed in this section because, we have now included the effect of the end cell geometry in the calculation. A summary of all important RF parameters calculated for the 5-cell cavity is presented in table 4.

**4. HIGHER ORDER MODE STUDIES**

So far we have discussed about the geometry optimization of the cavity from the point of view of better RF performance in the fundamental operating mode. A cylindrically symmetric RF cavity can support, along with the fundamental $TM_{010}$ mode resonating at the lowest frequency that is responsible for beam acceleration, higher order transverse electric (TE, having no electric field component parallel to the axis) as well as transverse magnetic (TM, having no magnetic field component parallel to axis) modes satisfying the appropriate boundary conditions. Except for trapped modes, a number of such HOMs having frequency less than the cut-off frequency of the beam pipe is possible in a cavity. The performance of an accelerating cavity is limited by the instabilities generated by these HOMs induced in the cavitydue to interaction of the beam with the cavity [1, 13]. While traversing through the cavity, the beam can excite the HOMs, which can interact again to the trailing part of the same bunch or the bunches behind it. Amongst these modes, mainly the monopole and dipole modes can influence a well collimated on-axis beam significantly. There exists a threshold beam current beyond which the instability is excited and the beam can not be transmitted through the accelerating structure. In superconducting cavities,this problem becomes crucial because of the high value of the quality factor $Q$ as well as the complex cavity shape. The high value of $Q$ results in the long decay time of the HOMs excited by the beam. The HOM couplers that couple out the HOMs from the cavity are generally put on the beam pipe only. Therefore, one has to perform HOM study of the cavity and make sure that amongst the prominent HOMs, there are no trapped modes and all the HOMs can be out-coupled. Also, the threshold current to excite the beam instability for each mode should be calculated and one needs to make sure that threshold current is higher than the desired beam current. For all these calculations, one needs to know the resonant frequencies and strengths of the HOMs supported in the cavity, which we calculate and report in this scetion.

We have done the *R*/*Q* calculations for both monopole and dipole modes, using SLANS. For higher order monopole modes, the mode strength is defined in the same manner as in the case of fundamental monopole mode and is indicative of kick imparted to the particle in longitudinal direction for a given amount of power loss on cavity walls, and is given by *R*/*Q*. For the dipole modes, the particles get a deflecting kick in the transverse direction and the figure of merit here is given by $R_\perp/Q$, which is expressed as

$$\frac{R_\perp}{Q} = \frac{1}{\omega_n U_n} \left| \int_{z_s}^{z_e} \left(\frac{\partial E_z}{\partial r}\right) e^{i\omega_n z/\beta c} dz \right|^2. \tag{5}$$

Note that the above expression is same as $k_n^2$ times the expression described in Ref [13] and in agreement with the commonly followed convention. Here, $k_n$, $\omega_n$ and $U_n$ denote wave vector, the angular frequency and the stored energy respectively of the $n^{th}$ mode. Here $\partial E_z/\partial r$ is the transverse gradient of the axial electric field. We have performed the calculations of *R*/*Q* and $R_\perp/Q$ for monopole and dipole modes respectively for the optimized cavity geometry described in this paper, where the radius of the beam pipe is kept as 50.00 mm. Since the cutoff frequency for this value of beam pipe radius is around 2 GHz, we fix the upper limit on the frequency for the HOM calculations as 2.5 GHz.

Figure 17 shows *R*/*Q* for monopole modes having frequency less than 2.5 GHz. We notice that the HOM at a frequency 1233.399 MHz has an *R/Q* ~ 124 Ω. Also, there exists three other HOMs at 1230.718 MHz, 2162.79 MHz and 1980.815 MHz frequencies respectively, having *R/Q* ~ 10 Ω. All the remaining monopole HOMs have *R/Q* less than 10 Ω.

Next, we calculate the $R_\perp/Q$ for dipole modes in figure18. We observe that there are only four higher order dipoles modes that are prominent, which appear at frequencies 945.36 MHz, 950.266 MHz, 1370.91 MHz and 1378.71MHz respectively and have their $R_\perp/Q$ values more than $10^4 \Omega/m^2$.

We use the above data to calculate the threshold current for the regenerative beam break up instability. For each dipole mode, there is a threshold current $I_{th}$ for this instability that is given by [23, 24]

$$I_{th} = \frac{\pi^3 \xi_0 k_n}{2 \cdot q (R_\perp/L_{cav}) \cdot L_{cav}^2}. \tag{6}$$

Here, $\xi_0$ is the beam energy and $L_{cav}$ is the length of the cavity. For the optimized cavity, $L_{cav} \sim 1$ m, and in the starting of the $\beta$=0.9 section, the particles have ~ 500 MeV kinetic energy. Note that the above formula is for a cw beam. In table 5, we have presented the prominent HOMs and their corresponding threshold currents. Note that the lowest value of the threshold current for the excitation of this instability is 0.85 mA for the HOM excited at 1370.91 MHz. For the ISNS project, it is being planned that the beam pulse current will be 4 mA with a pulse width of 2 ms and rep. rate of 50 Hz. The average cw beam current will be 0.4 mA in this case, which is significantly less than the threshold beam current of 0.85 mA for the excitation of regenerative beam break up instability. Note that with the help of HOM couplers, $R_\perp$ can be further reduced by lowering the value of $Q$ and this will further enhance the threshold beam current.

Another important effect of the beam cavity interaction is manifested in terms of wake fields, which affect the longitudinal as well as the transverse beam dynamics in the cavity. We have studied the loss factors as well as the longitudinal and transverse wake potentials using the code ABCI. Figure 19 shows the results of this calculation for the cavity geometry described in the paper for a Gaussian bunch having rms bunch length $\sigma_z$ of 5.00 mm and 1pC charge. We find that the longitudinal loss factor $k_\parallel$ for this case is equal to 2.88 V/pC and the transverse loss factor $k_\perp$ is 11.55 V/pC/m. The variation of $k_\parallel$ and $k_\perp$ as a function of $\sigma_z$ is shown in figure 20.

Calculations of wake potentials and loss factors have been done primarily for beam dynamics calculations, which will be taken up later. However, we can get a quick estimate of effects of wakefield using the plots we have described. Two important implications of wakefields are – (i) energy spread in the beam induced by the wake field, and (ii) parasitic heat loss on cavity walls due to excitation of higher order modes. First, we discuss the calculation to estimate the energy spread $\delta\xi$ induced in a single bunch due to wakefield, which is given by [13]

$$\delta\xi = -e \times q \times W_z^{\prime} \times s. \qquad (7)$$

Here, $W_z^{\prime}$ is the slope of the longitudinal wake potential, which is estimated as as ~ 0.318 V/pC/mm from figure 19, and $s$ is the bunch length, which is approximately 22 mm. In our case, inside the 2 ms long macropulse having a 4 mA of average current, the micropulses repeat at a frequency of 325 MHz. Thus charge $q$ per micropulse is ~ 12.3 pC. Using these numbers, Eq. (7) gives us $\delta\xi$ = 86 keV. The energy gain form a single cavity is 18.5 MeV. Therefore, the relative spread in the beam energy will be ~ 0.46%. The reported value for the beam energy spread in the TESLA cavity for a Gaussian beam bunch of ~ 1 mm length is 0.2 %. As described in Ref. [25], by proper adjustment of input beam phase, it is possible to minimize the energy spread induced by the wakefield. Next, we discuss the parasitic heat loss on the cavity walls due to the excitation of wakefields by the beam. The power $P_0$ appearing as parasitic heat loss is related to the longitudinal loss factor $k_{//}$ by the following equation [8]

$$P_0 = (PRR) \times q^2 \times k_{\parallel}, \qquad (8)$$

where PRR is the pulse repetition rate. In our case, taking PRR as 325 MHz, $q$ = 12.31 pC and $k_{//}$ = 2.88 V/pC, we get power dissipation within the macropulse as 142 mW and the CW average will be 14. 2 mW. We would like to point out that the above calculation is under the assumption that in a macropulse, wakefields from different micropulses are adding incoherently. This could be a valid approximation since the frequency of significant HOM supported in the cavity is not an integral multiple of PRR.

**5. LORENTZ FORCE DETUNING STUDIES**

Electromagnetic field in an RF cavity induces surface current and surface charges on the wall of the cavity, which experience the Lorentz force due to interaction with the electromagnetic field. This generates a pressure on the cavity surface given by [2,16],

$$P = \frac{1}{4}(\varepsilon_0 E^2 - \mu_0 H^2). \tag{9}$$

As a consequence of this pressure, the cavity geometry gets deformed and its resonant frequency shifts, which is known as the Lorentz force detuning (LFD). Because of the high value of $Q$ of superconducting cavities, the bandwidth is small and hence, even a small detuning may result in a large reflection of the input power from the cavity. For example, in our design, the loaded $Q$ is expected to be around $5 \times 10^6$ assuming the beam pulse structure described in the last section. For the operating frequency of 650 MHz, this gives us the bandwidth of the loaded cavity as 130 Hz, which is very small. Hence, to reduce the value of this detuning, along with the proper selection of the wall thickness of the cavity, additional mechanical constraint in the form of stiffener ring [26] is introduced in between the consecutive cells of the cavity. In addition, stiffener ring is also placed between the end cell and the wall of the helium vessel. The reduction of the detuning strongly depends on the stiffness of the helium vessel. We would like to point out here that the helium vessel not only influences the tuning efficiency but it influences the tuner also.

There are two types of scenarios that are possible – first, for a cw beam, as it is evident from the above mentioned expression, the time averaged radiation pressure will yield a steady shift in frequency. Therefore, this is known as the static Lorentz force detuning. To get rid of this, one can stretch the cavity in the longitudinal direction in order to compensate for this negative shift in the frequency. Second, for a pulsed machine, there is a possibility of an unwanted coupling between the resonant frequency for the mechanical oscillations of the structure and the repetition rate of the pulse. This results in a resonance popularly known as dynamic resonance. This is called the dynamic LFD [27], which is dependent on the RF pulse structure. To control this dynamic behavior, piezo tuners are used.

In this section, we first report the calculations of static LFD that we obtained from the simulations done using ANSYS. We have studied the LFD as a function of the stiffness of helium vessel as well as the location of the stiffener ring. In order to compensate for the LFD, we have also studied the required range for tuning the resonant frequency by elongating the cavity. For our calculations, we have considered a simplistic geometry of the vessel and repeated the above mentioned simulations for different values of stiffness of the helium vessel. A helium vessel of 504 mm diameter and having the end closure in "tori-flat" shape with a torus radius of 35 mm has been modeled for this study, which is shown in figure 21. The term "tori" here indicates that edge of the end closure of the helium vessel is actually a section of a toroid. For the tori-flat geometry, the end closure of the cylindrical helium vessel is flat and has a toroidal shape near the edge. On the other hand, For the tori-spherical case, the end closure has spherical shape and has a toroidal shape near the edge. The wall thickness of the helium vessel is taken as 4 mm and the material is taken as titanium. The cavity wall thickness is also taken as 4 mm. For these calculations, subroutines have been written in ANSYS parametric design language.

The stiffness of the helium vessel was increased to gauge its effect. The locations of stiffeners between the end closure of the helium vessel and the end cell, as well as between the individual cells play an important role in restricting the LFD. In addition, the stiffeners also affect the ability to tune the cavity, which we have studied. The locations of stiffeners are shown in figure 21. Pressure on the cavity walls due to Lorentz force in the presence of stiffeners is shown in figure 22. Figure 23 shows the LFD calculated as a function of the stiffness of the helium vessel, in the absence of stiffener ring. We find that the LFD decreases significantly when we increase the stiffness from 2.3 kN/mm to 6.5 kN/mm and its absolute value decreases from 2.6 kHz to 1.1 kHz. We have checked that this range of stiffness is practically achievable. This can be done by increasing the thickness of the helium vessel (up to 6 mm), and by changing the shape of the end closure from "tori-flat" to "tori-spherical". Next, we discuss the reduction in LFD that can be achieved with the help of stiffener rings. Figure 24 shows the LFD as a function of the radial location of the stiffener ring, taking the stiffness of the helium vessel as 4.9 kN/mm. The radial location of the stiffeners was kept same for all the cells. We find that the placement of stiffeners towards the equator region leads to a substantial reduction in the LFD, as seen in figure 25.

We now study the possibility of compensating the LFD using a tuner. We found that the static LFD can be compensated by elongating the cavity, for which helium vessel needs to be displaced by 4.75 μm, and the preferred location of the stiffener ring that gives full compensation of LFD is seen in figure 25. These calculations are presented for operation at acceleration gradient of 18.5 MV/m. We notice that the preferred radial location of stiffener ring is on the common tangent region of elliptic cells, where it is relatively easier to weld the stiffener ring. We conclude from figure 24 and figure25 that the location of stiffener ring should be decided keeping in consideration the ability to compensate for the LFD by displacement of the helium vessel. For example, if the stiffener ring is located near the equator, the static LFD will be small, but the cavity cannot be tuned. Hence, the displacement required to be given to the helium vessel to compensate for the LFD may be large. This may not be easy to implement. This is the reason why we have chosen the common tangent region of elliptic cells to locate the stiffener ring.

## 6. DISCUSSIONS AND CONCLUSIONS

In this paper, we have presented the design of a $\beta_g$=0.9, 650 MHz SRF cavity for the proposed ISNS project. The geometry of the mid cell is described by seven independent parameters. The optimized value of these parameters is chosen using the following general procedure, which is summarized below:

- First, the iris radius $R_{iris}$ is decided on the basis of beam dynamics considerations and the requirement of the threshold current for beam instability. The minimum possible value allowed with these considerations is suitably chosen and fixed.
- The half cell length is chosen as $L = \beta_g \lambda / 4$.
- For each design simulation, $R_{eq}$ is tuned to achieve the design frequency.
- The maximum possible value of $\alpha$ that is allowed by practical considerations is chosen.
- Out of the seven independent parameter that define the mid cell geometry, we are now left with only three independent parameters. Initially we keep two additional constraints, i.e., $a = b$ and $A = B$. We are thus left with only one independent variable. We plot $B_{pk}/E_{acc}$ as well as $E_{pk}/E_{acc}$ as a function of $B$. We find that $B_{pk}/E_{acc}$ monotonically decreases with $B$, whereas $E_{pk}/E_{acc}$ initially remains nearly constant and then starts increasing rapidly. We choose and fix the value of $B$, where $E_{pk}/E_{acc}$ just starts increasing rapidly.
- After having fixed the value of $B$, we are now left with only two independent variables. We plot $B_{pk}/E_{acc}$ as well as $E_{pk}/E_{acc}$ as a function of $A$ for different values of $a/b$. We fix a target value of $E_{pk}/E_{acc}$ and using these plots, find out the maximum possible value of $A$ with which it is possible to obtain this target value for $E_{pk}/E_{acc}$. Around this value of $A$, we plot $E_{pk}/E_{acc}$ as a function of the ratio $a/b$ for different values of $A$. we find out the value of $a/b$ that gives the minimum value of $E_{pk}/E_{acc}$. We thus have an approximate value of $A$ and $a/b$ around which further fine tuning of the optimization can be done.
- Finally, using the target value of $E_{pk}/E_{acc}$ as a constraint, we plot $B_{pk}/E_{acc}$ as a function of $a/b$ as well as $A$. We find out the value of $a/b$ and $A$ that minimizes $B_{pk}/E_{acc}$ and these are taken as final design values. We thus have all the seven parameters that define the geometry of the mid cell.

The procedure that we have summarized above is a new and generalized logical approach towards the design of superconducting elliptic cavity. As we have mentioned earlier in the paper, the typical approach followed by the other authors in the literature for this purpose, is the multi-dimensional optimization as described in Refs. 5 and 9. Here, we have instead given a step by step, one-dimensional optimization technique to ultimately perform a multi-variable optimization, and more importantly, at every step, we have provided the physics arguments to understand the particular dependence and the behavior. Our approach is unique in this sense compared to the existing approach used in the design of superconducting elliptic cavity. Using this approach, it is possible to do the cavity design starting from scratch.

After having completed the optimization of the mid cell geometry, the design of the end cell of the cavity is optimized such that the resonant frequency of the multi cell cavity matches with that of the mid cells. This criterion leads to maximum field flatness in the cavity.

For the optimized design of the multi cell cavity, we have estimated the shunt impedance of higher monopole as well as dipole modes. We have checked that amongst the prominent HOMs, there are no trapped modes. Using the shunt impedance data of dipole HOMs, we have estimated the threshold current for the regenerative beam break up instability. Since the threshold current is higher than the design current, our choice of $R_{iris}$ is acceptable. We would like to emphasize that the number of cells in the cavity also affects the threshold for regenerative beam break up instability as seen from Eq. (6). As the number of cells increases, the threshold current decreases.

Finally, we have also presented the LFD studies and have shown that keeping the thickness of cavity wall as 4 mm and choosing appropriate position of the stiffener ring and appropriate value of the stiffness of the helium vessel, it is possible to tune the cavity to compensate for the LFD. Studies on dynamic LFD will be taken up in future.

We would like to emphasize that although we have used standard software like, SUPERFISH, SLANS and ANSYS, we have written independent computer programs to utilize these codes in an efficient manner to perform the detailed optimization calculations described here. For example, SUPERFISH has been used in the step by step optimization procedure, for which a separate wrapper code in C programming language has been written to speed up the optimization simulations. For the LFD calculations, separate program has been written in ANSYS parametric language and a module has been developed that enables us to integrate the details about helium vessel stiffness and stiffener location to the electromagnetic calculation of the Lorentz pressure from electromagnetic fields. The resulting structural deformations and change in the resonating frequency is studied in a single software environment promoting an efficient and accurate analysis. This would open up new future possibilities of detailed study of the effect of cavity thickness variation, and also the dynamic Lorentz Force detuning calculations in the same model.

We have also compared our design with that of Project-X design of Fermi National Accelerator Laboratory, USA, which is also for the RF frequency 650 MHz. We find that our optimized cavity geometry for the mid cell is similar to that reported for Project-X in Ref. [28]. The basis for choosing the optimized parameters for the mid cell geometry is however not described in Ref. [28]. Here, we have given the details of optimization study to arrive at the geometry of the mid cell. The frequencies and shunt impedances of HOMs reported here are in agreement with Ref. [29]. Our studies on LFD should be applicable to Project-X design also.

To summarize, we have presented basic design of a $\beta_g$=0.9, 650 MHz SRF cavity and discussed the basic design approach. The beam dynamics studies using an array of such cavities and focusing lattice in between them will be taken up in near future.

**Acknowledgement**

It is a pleasure to thank Dr. P. D. Gupta for constant encouragement, and Dr. S. B. Roy for helpful discussions and suggestions on the manuscript.

**Tables;**

**Table 1.** Optimized parameters for the mid cell geometry.

| Parameter | Magnitude | Unit |
|---|---|---|
| $R_{iris}$ | 50.00 | mm. |
| $R_{eq}$ | 199.925 | mm. |
| $L$ | 103.774 | mm. |
| $A$ | 83.275 | mm. |
| $B$ | 84.00 | mm. |
| $a$ | 16.788 | mm. |
| $b$ | 29.453 | mm. |
| $\alpha$ | $85^0$ | |

**Table 2.** RF parameters of the optimized mid cell geometry for the π mode operation.

| RF Parameter | Magnitude | Unit |
|---|---|---|
| Frequency | 650 | MHz. |
| Transit-time factor(T) | 0.757 | |
| $E_{acc}$ | 18.6 | MV/m. |
| $Q_0$ | $> 1 \times 10^{10}$ | |
| $R/Q_0$ | 60.8 | Ω. |
| $G$ | 257.2 | Ω. |
| $E_{pk}$ | 37.2 | MV/m. |
| $B_{pk}$ | 69.62 | mT |

**Table 3.** Calculation for the cell to cell coupling coefficient

| Mode | Resonant Frequency(MHz.) | Cell to cell coupling coefficient |
|---|---|---|
| $\pi$ | $f_\pi$ = 650.00082 MHz. | $k_c = 2 \times \dfrac{(f_\pi - f_0)}{(f_\pi + f_0)}\% = 0.803\%$ |
| 0 | $f_\pi$ = 644.80535 MHz. | |

**Table 4.** RF parameters of the optimized $\beta_g$ = 0.9, 5-cell, 650 MHz cavity for $\pi$ mode operation.

| RF parameters | Magnitudes | Units |
|---|---|---|
| Frequency | 650 | MHZ. |
| Transit-time factor($T$) | 0.717 | |
| Acc. Gradient ($E_0T$) | 18.6 | MV/m. |
| $R/Q$ | 608.7 | $\Omega$. |
| $G$ | 257.6 | $\Omega$. |
| $E_{pk}$ | 37.82 | MV/m. |
| $B_{pk}$ | 69.831 | mT |

**Table 5.** Details of the prominent dipole modes supported by the cavity.

| Mode frequency $f$ (in MHz) | $Q$ | $R_\perp/Q$ ($\Omega/m^2$) | $I_{th}$ (in mA) |
|---|---|---|---|
| 945.36 | $5.143 \times 10^9$ | 13751.68 | 4.237 |
| 950.266 | $3.185 \times 10^9$ | 33461.56 | 2.827 |
| 1370.91 | $8.495 \times 10^9$ | 60083.52 | 0.851 |
| 1378.71 | $7.466 \times 10^9$ | 17161.29 | 3.411 |

**Figures;**

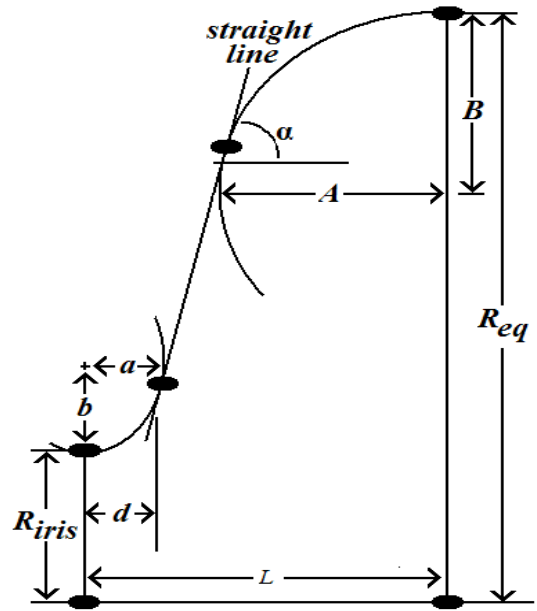

**Figure 1.** Schematic of the half-cell of an elliptic cavity.

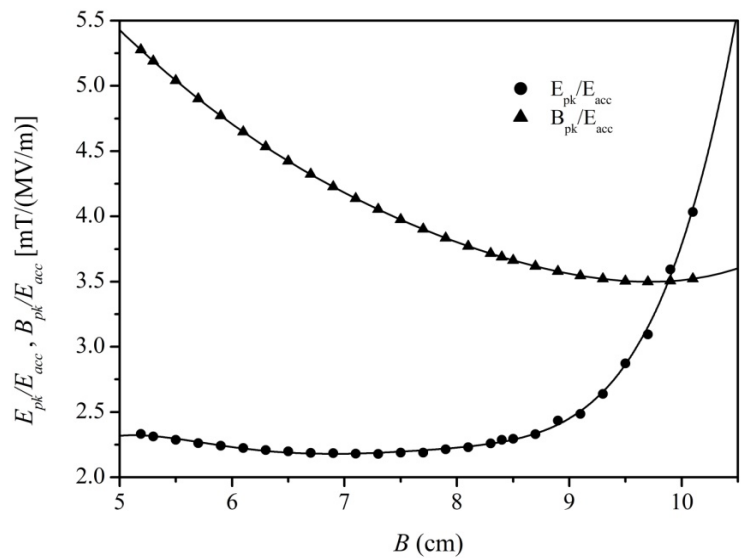

**Figure 2.** Variation of $B_{pk}/E_{acc}$ and $E_{pk}/E_{acc}$ as a function of $B$. Here, $a/b = A/B = 1$, and $\alpha = 90^0$.

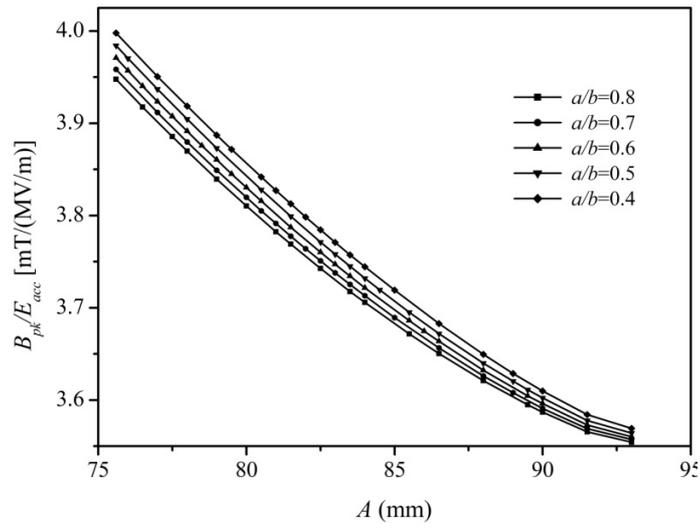

**Figure 3.** Plots of $B_{pk}/E_{acc}$ as a function of $A$ for different values of $a/b$. Here, $B = 84$ mm, and $\alpha = 85^0$.

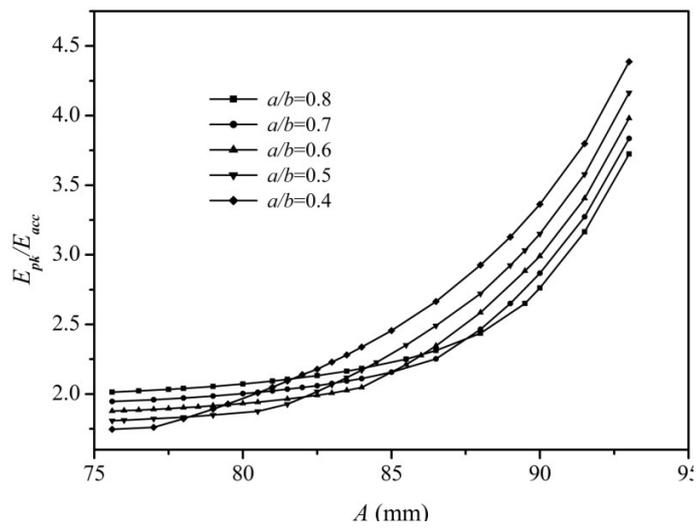

**Figure 4.** Plot of $E_{pk}/E_{acc}$ as a function of $A$ for different values of $a/b$. Here, $B = 84$ mm and $\alpha = 85^0$.

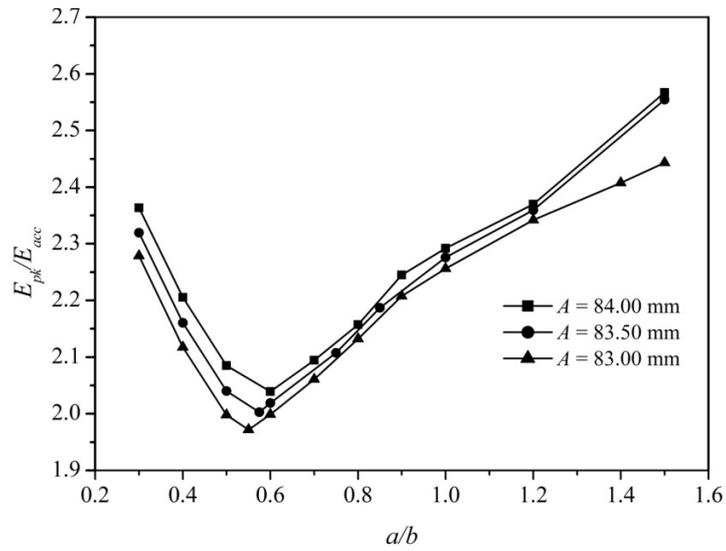

**Figure 5.** Plot of $E_{pk}/E_{acc}$ as a function of $a/b$ for different values of $A$.

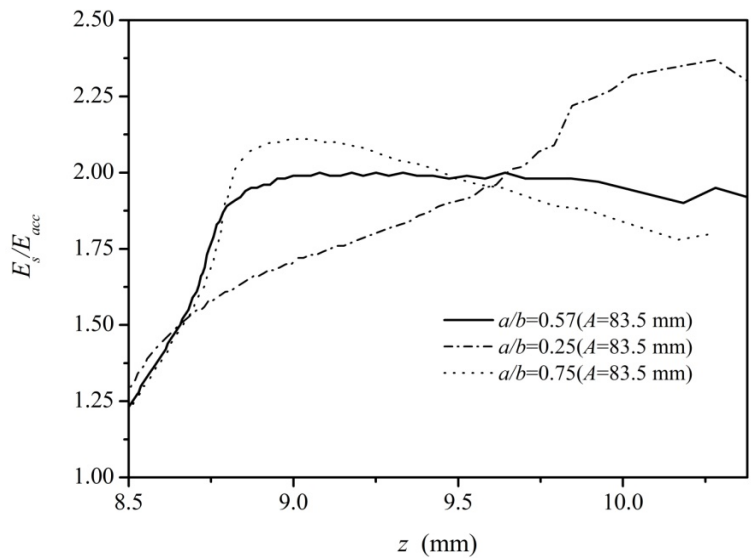

**Figure 6.** Plot of $E_s/E_{acc}$ as a function of $z$. Here, the solid line corresponds to $a/b = 0.57$, for which $E_{pk}/E_{acc}$ shows a minima in Fig. 5. The other two curves are for $a/b = 0.25$ and $0.75$ respectively, which show local maxima. Here $z = 0$ corresponds to the equator plane.

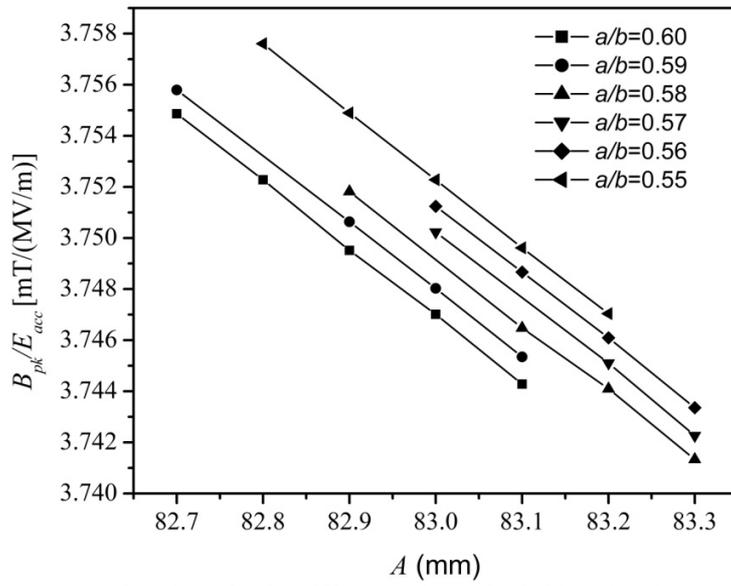

**Figure 7.** Plot of $B_{pk}/E_{acc}$ as a function of $A$ for different values of $a/b$. Note that the range in $A$ in this plot is narrowed in comparison to Fig. 3.

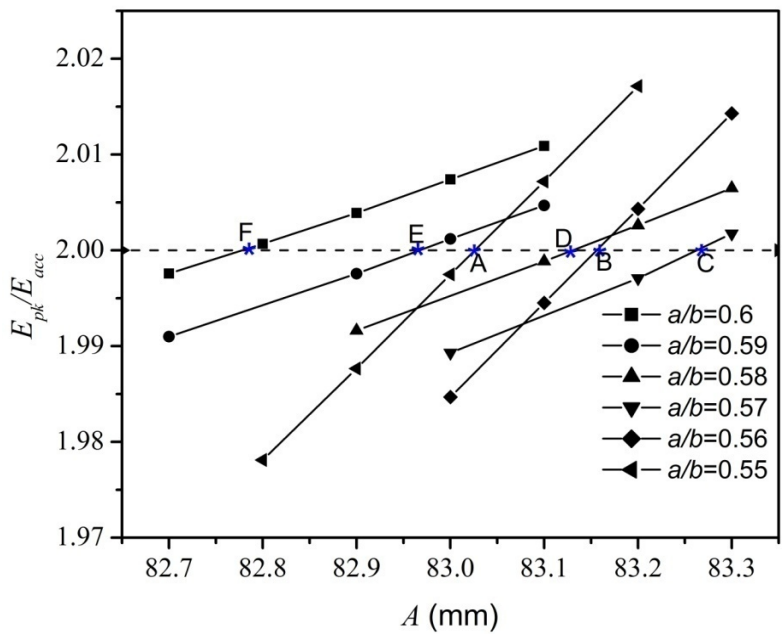

**Figure 8.** Plot of $E_{pk}/E_{acc}$ as a function of $A$ for different values of $a/b$. Note that the range in $A$ in this plot is narrowed in comparison to Fig. 4.

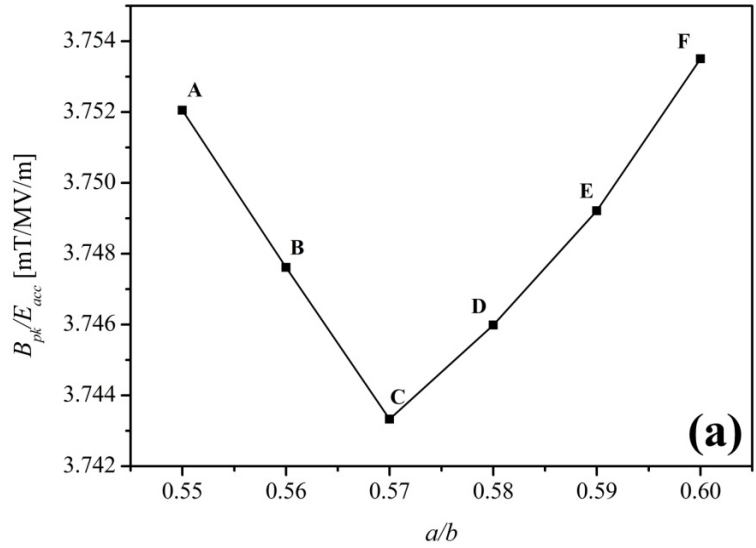

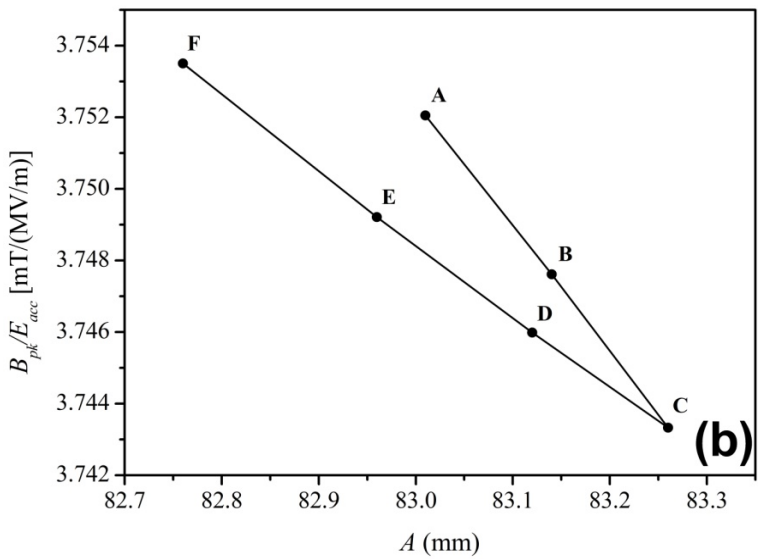

**Figure 9.** Plot of $B_{pk}/E_{acc}$ as a function of $a/b$ (a) and as a function of $A$ (b). For each of these points, $E_{pk}/E_{acc} = 2.0$. Data for plotting these curves is taken from Figs. 7 and 8.

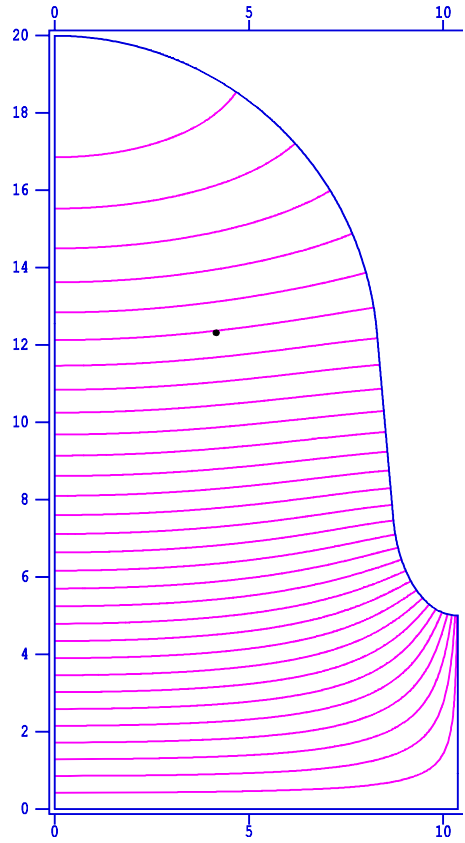

**Figure 10.** Field contours for the optimized mid cell geometry resonating in π mode.

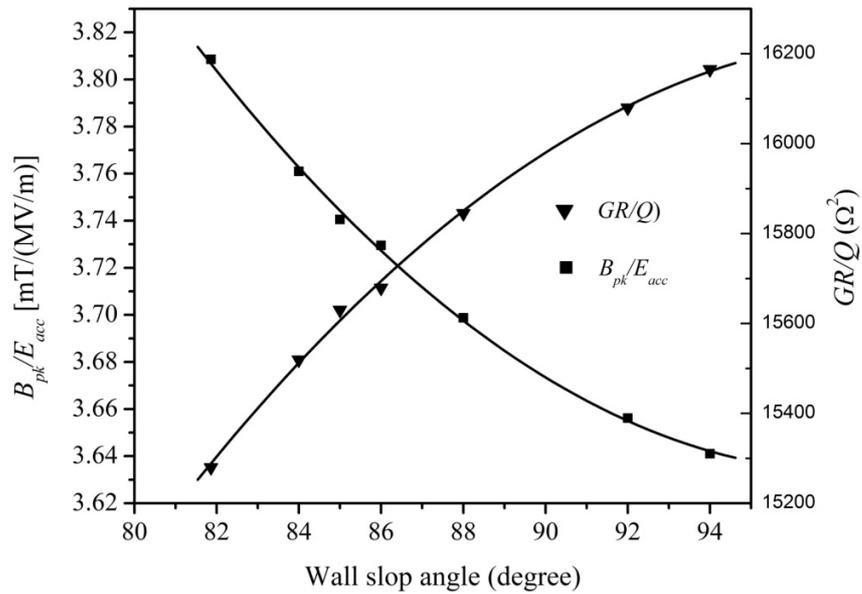

**Figure 11.** Optimum value of $B_{pk}/E_{acc}$ as a function of the wall angle $\alpha$. Here the value of $E_{pk}/E_{acc}$ is equal to 2 for each data point.

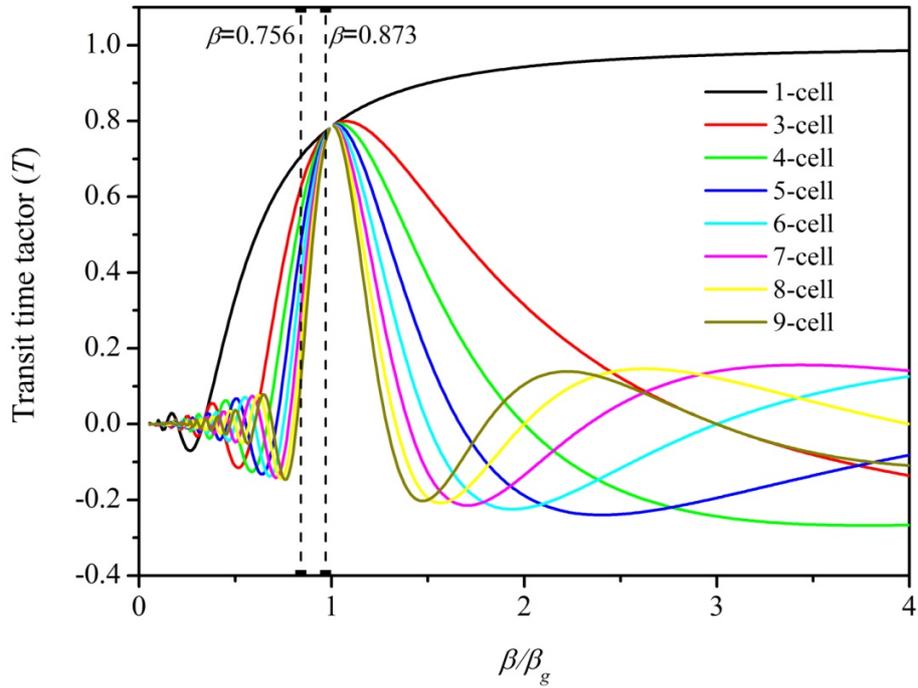

**Figure12.** Variation of the transit time factor $T$ as a function of the normalized particle velocity ($\beta/\beta_g$). The two vertical lines correspond to $\beta/\beta_g = 0.84$ and 0.97, as explained in the text.

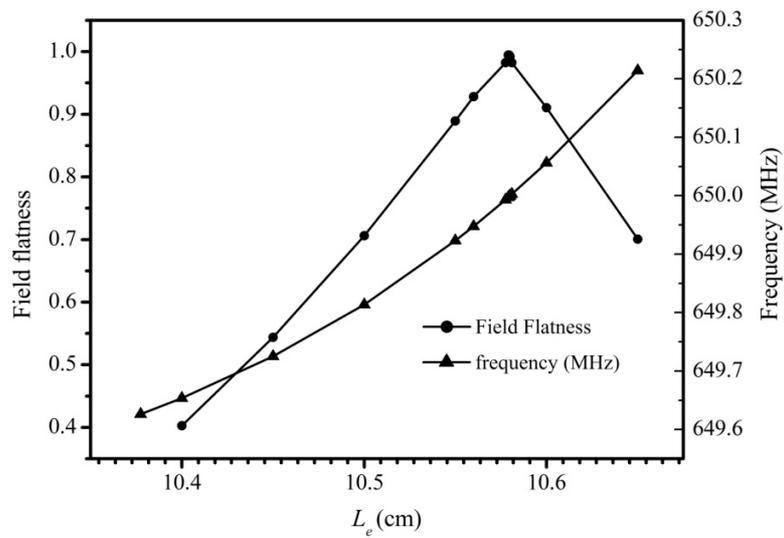

**Figure13.** Field flatness and resonant frequency as a function of the length $L_e$ of the end half cell.

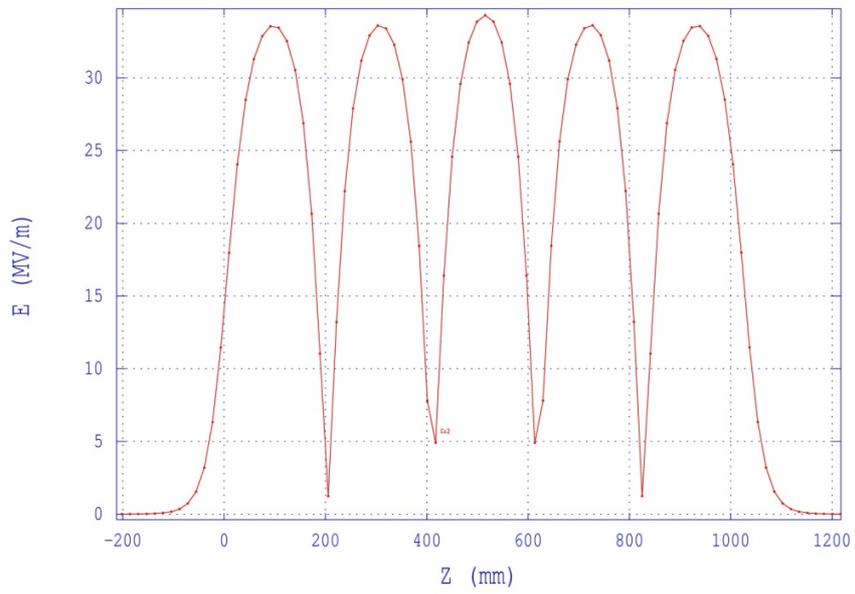

**Figure 14.** Variation of the magntidue of the axial electric field along the beam axis for the optimized 5-cell cavity.

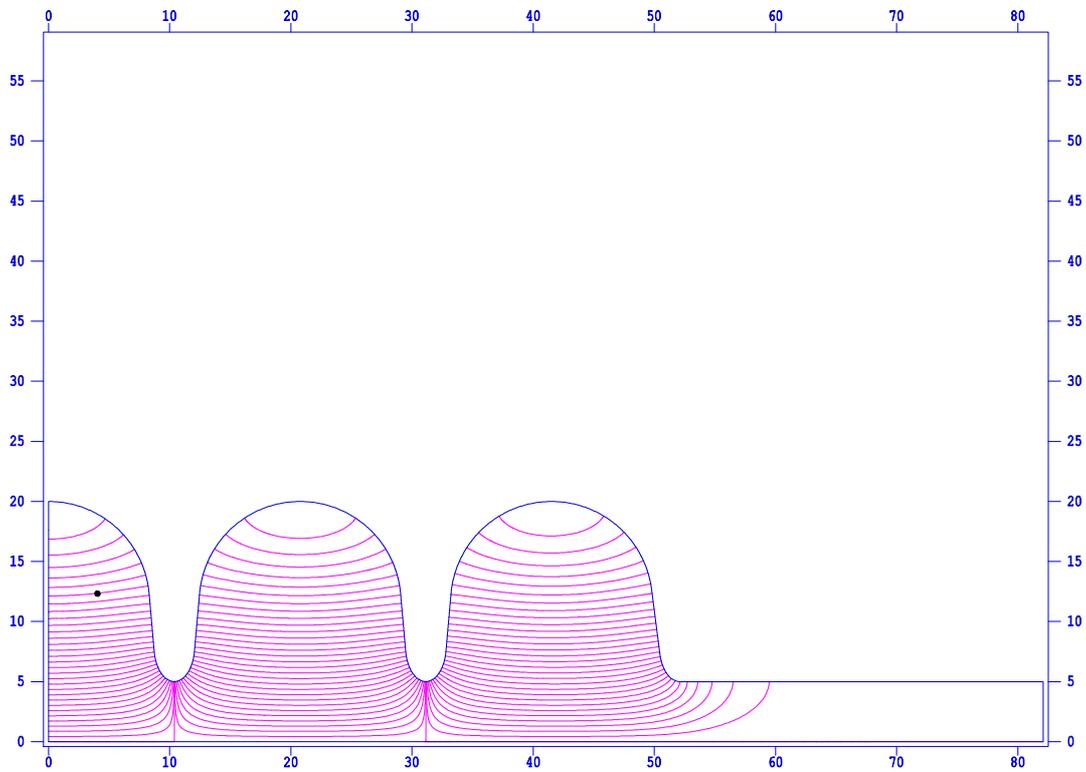

**Figure 15.** Field contours for one half of the 5-cell cavity.

38 | P a g e

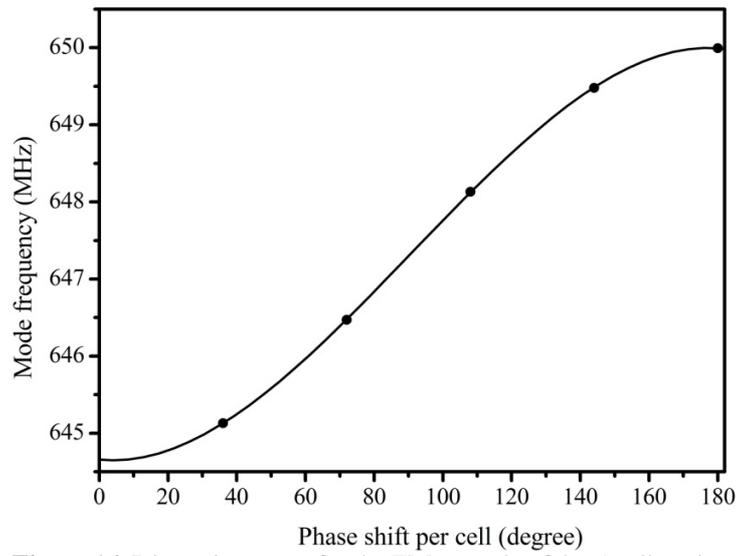

**Figure 16.** Dispersion curve for the TM$_{010}$ mode of the 5-cell cavity.

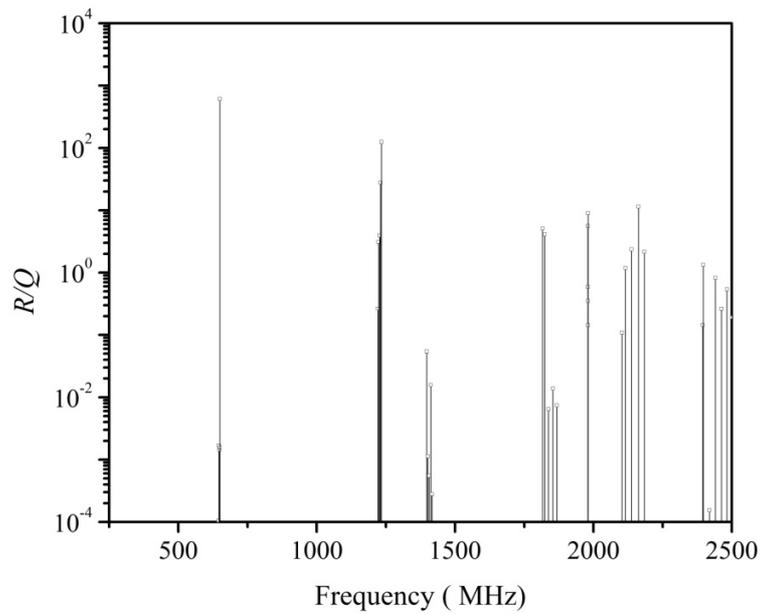

**Figure 17.** *R*/*Q* of the monopole modes plotted against the corresponding frequencies.

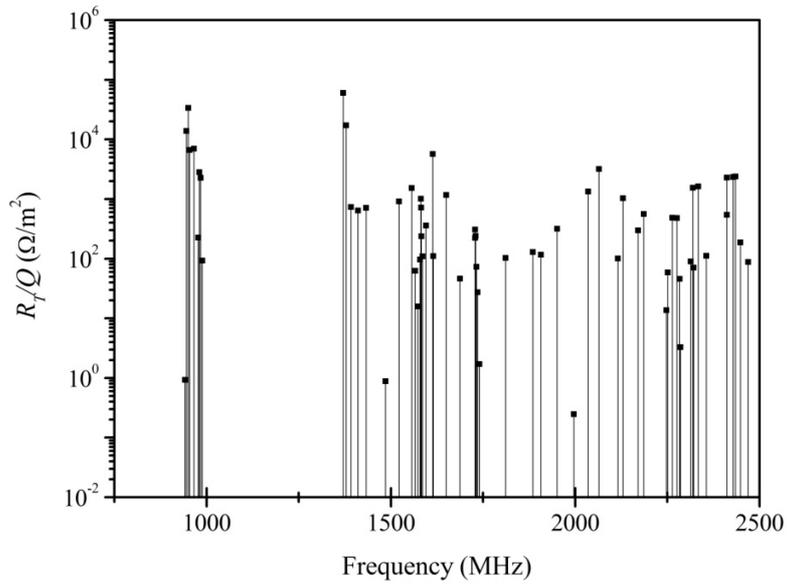

**Figure 18.** $R_T/Q$ of the dipole modes plotted against corresponding frequencies.

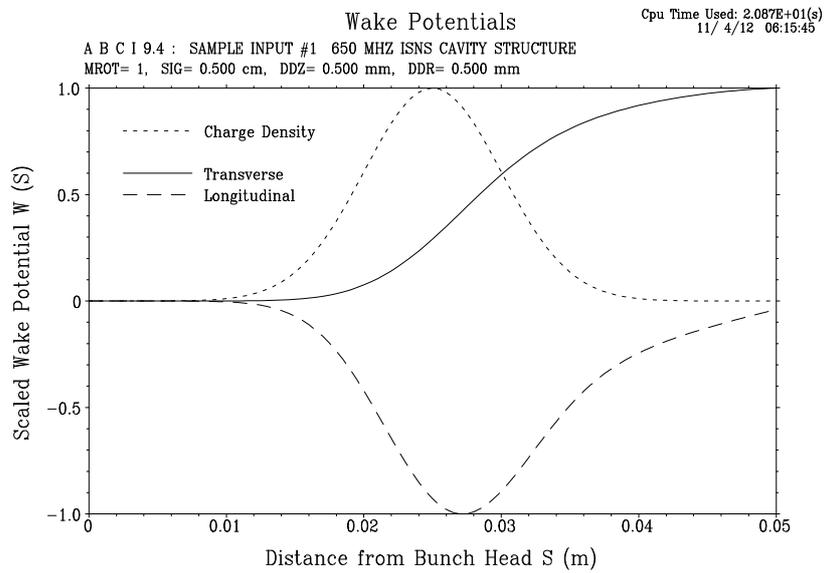

**Figure 19.** Longitudinal and transverse wake potential generated by a Gaussian bunch having 1 pC charge passing through the optimized 5-cell cavity.

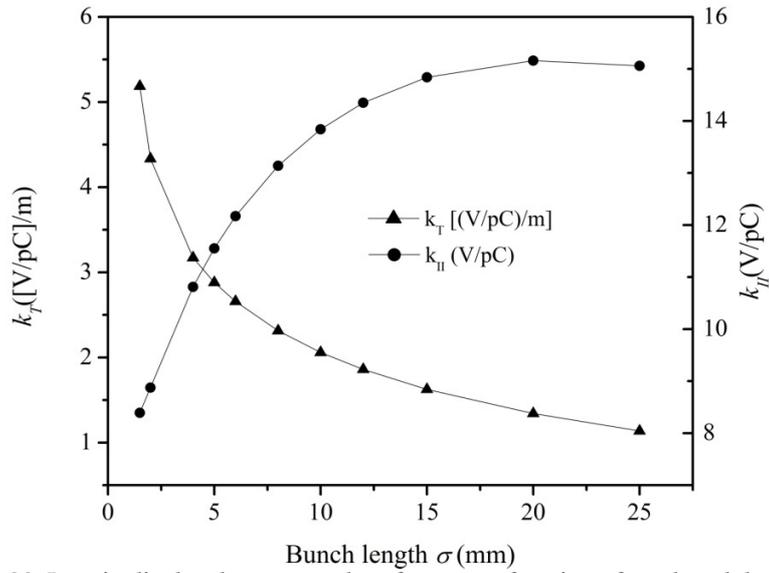

**Figure 20.** Longitudinal and transverse loss factor as a function of rms bunch length σ_z.

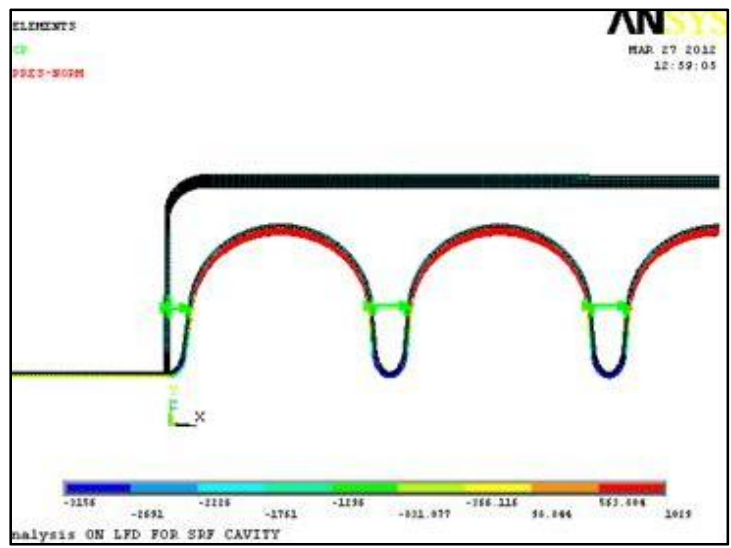

**Figure 21.** Schematic of the multi cell cavity along with the helium vessel. Green lines connecting the cells are the stiffeners.

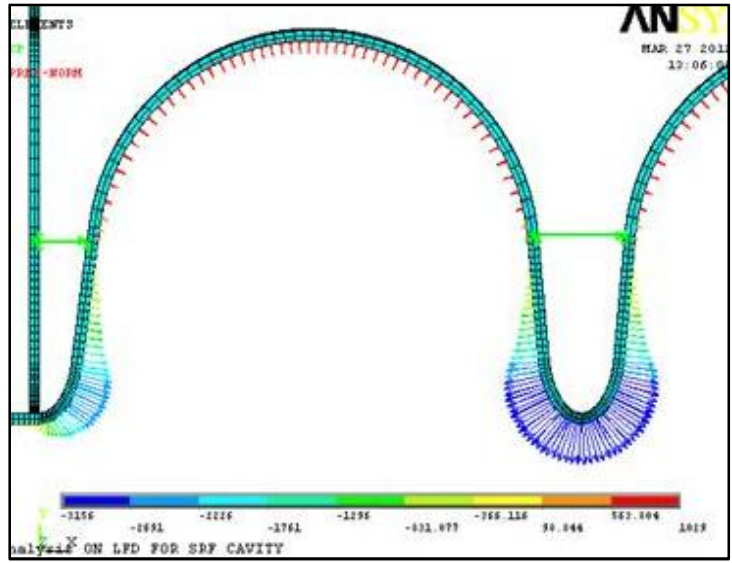

**Figure 22.** Arrows showing the direction of the Lorentz force at different locations in the cavity.

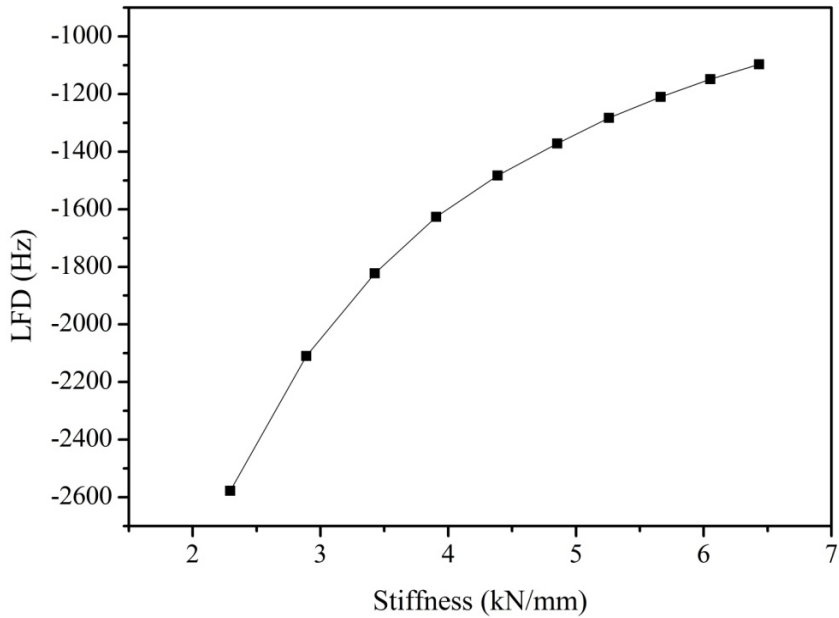

**Figure23.** LFD as a function of stiffness of the helium vessel. These calculation are for the case when there is no stiffener ring.

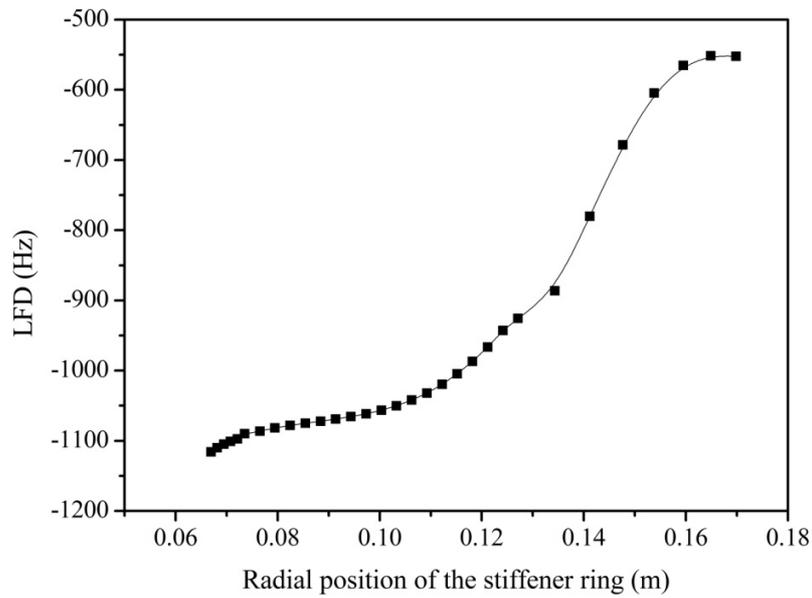

**Figure 24.** LFD as a function of the radial position of the stiffener rings. Stiffness of the helium vessel is taken as 4.9kN/mm in these calculations.

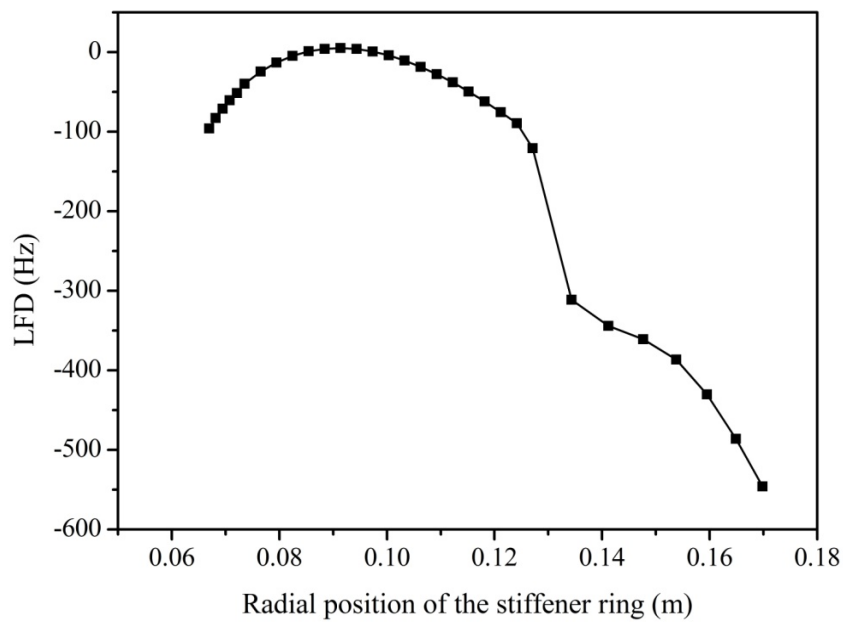

**Figure 25.** LFD as a function of the radial position of the stiffener rings. Stiffness of the helium vessel is taken as 4.9kN/mm. The compensation due to cavity elongation provided by a displacement of 4.75 µm of the helium vessel is taken into account here.